\documentstyle[aaspp4,12pt]{article}

\lefthead{J.J.M. in 't Zand, A. Baykal \& T.E. Strohmayer}
\righthead{4U~1907+09}

\begin{document}

\title{RECENT X-RAY MEASUREMENTS OF THE ACCRETION-POWERED PULSAR 4U~1907+09}
\author{J.J.M.~in~'t~Zand}
\affil{Space Research Organization Netherlands, Sorbonnelaan 2, 3584 CA Utrecht, 
       the Netherlands; jeanz@sron.ruu.nl}
\author{A.~Baykal}
\affil{Physics Department, Middle East Technical University, Ankara 06531, Turkey; 
	altan@astroa.physics.metu.edu.tr}
\author{T.E.~Strohmayer\altaffilmark{1}}
\affil{Laboratory for High-Energy Astronomy, NASA - Goddard Space Flight 
       Center, Greenbelt, MD 20771, U.S.A.; stroh@lheamail.gsfc.nasa.gov}
\altaffilmark{1}{USRA research scientist}

\noindent
\begin{center}
{\it To appear in Astrophysical Journal, Vol. 496, on March 20, 1998}
\end{center}

\begin{abstract}
X-ray observations of the accreting X-ray pulsar 4U~1907+09, obtained during February
1996 with the {\em Proportional Counter Array\/} on the {\em Rossi X-ray Timing 
Experiment\/} (RXTE), have enabled the first measurement of the intrinsic pulse period 
$P_{\rm pulse}$ since 1984: $P_{\rm pulse}=440.341^{+0.012}_{-0.017}$~s. 
4U~1907+09 is in a binary system with a blue supergiant. The orbital parameters were
solved and this enabled the correction for orbital delay effects of a measurement 
of $P_{\rm pulse}$ obtained in 1990 with Ginga.
Thus, three spin down rates could be extracted from four pulse periods 
obtained in 1983, 1984, 1990, and 1996. These are within 8\% equal to a value 
of $\dot{P}_{\rm pulse}=$+0.225~s~yr$^{-1}$. This suggest that the pulsar is 
perhaps in a monotonous spin down mode since its discovery in 1983.
Furthermore, the RXTE observations show transient $\sim$18~s oscillations during 
a flare that lasted about 1 hour. The oscillations may be interpreted as 
Keplerian motion of an accretion disk near the magnetospheric radius. This,
and the notion that the co-rotation radius is much larger than any conceivable 
value for the magnetospheric radius (because of the long spin period), 
renders it unlikely that this pulsar spins near equilibrium like is suspected 
for other slowing accreting X-ray pulsars. We suggest as an alternative 
that perhaps the frequent occurrence of a retrograde transient accretion 
disk may be consistently slowing the pulsar down. Further observations of
flares can provide more evidence of this.
\end{abstract}



\section{Introduction}
4U~1907+09 is an X-ray pulsar powered by accretion of wind material from a blue 
supergiant companion star. It was discovered as an X-ray source by
Giacconi et~al. (1971) and has been studied using instruments onboard 
Ariel V (Marshall \& Ricketts 1980, MR80), Tenma (Makishima et~al. 1984, M84),
EXOSAT (Cook \& Page 1987, CP87), and Ginga (Makishima \& Mihara 1992,
Mihara 1995). MR80 first determined the orbital period of the binary at 
8.38~days through an analysis of data taken over the course of 5 years 
(between 1974 and 1980) from a survey 
instrument on Ariel V with a net exposure time of about $\frac{1}{2}$~year. 
A folded lightcurve of these data shows a pronounced primary flare and a 
dimmer and irregular secondary flare. M84 observed 4U~1907+09 with Tenma in 1983 
and discovered the pulsar with a pulse period of 437.5~s. Also, through a good 
time coverage of the binary orbit, they were able to confirm the occurrence 
of two phase-locked flares. The secondary flare in the Tenma data appears as
bright as the primary flare. CP87 discuss EXOSAT data with a small though
reasonably uniform coverage of the orbit. They also find evidence for a 
phase-locked primary and secondary flare. Through combining Tenma and EXOSAT 
data, they were able to determine the binary orbit most accurately and found 
an eccentricity of 0.16$^{+0.14}_{-0.11}$. A measurement of the pulse period 
revealed an average spin down rate of +0.23~s~yr$^{-1}$ since the Tenma 
measurement 270~d earlier. 4U~1907+09 was observed with Ginga in 1990.
Makishima \& Mihara (1992) report a cyclotron feature at 21~keV found during
these observations. Mihara (1995) measured
the pulse period using the Ginga data, without correcting for the binary orbit
its value was determined to be 439.47~s. Sadeh \& Livio (1982), using data from 
the HEAO A-1 instrument, report the occurrence of 15~ms oscillations during 4 
out of 20 scans of the source each lasting about 10~s. The oscillation period 
was seen to change during each scan.

In February 1996, 4U~1907+09 was observed with the narrow-field instruments on the 
{\it Rossi X-ray Timing Explorer\/} (RXTE). A previous paper (In~'t~Zand, 
Strohmayer \& Baykal, 1997, ISB97) reported a frequent occurrence of 
complete dips in the pulsar signal with durations between a few minutes and 
1$\frac{1}{2}$~hrs as found in these data. In the current paper we 
report about the same data but concentrate on variability on time scales 
shorter than or equal to the pulse period. We discuss the pulse period and its 
history since 1983 and present evidence for the occurrence of transient 
oscillations. 
These timing diagnostics enable one to address interesting questions about the 
presence of an accretion disk and the angular momentum transfer in the binary. 
Furthermore, we briefly discuss the general
spectral trend as a function of pulse phase.


\section{Observations}
\label{observations}

RXTE-PCA (Zhang et~al. 1993) consists of five identical proportional counters that
are co-aligned to the same point in the sky. Collimators limit the field of 
view (FOV) to 1 square degree. The total geometric collecting area is 
approximately 6250~cm$^2$, the effective sensitive photon energy range is
from 2 to 60~keV. The satellite is in a low-earth orbit so that
long observations may be broken up by earth occultations and passes through
the South Atlantic Anomaly.

RXTE-PCA observed 4U~1907+09 during 4 observation runs which are detailed in 
Table~\ref{tab1} and total 79.4~ks.
The mode of data collection for the four observation runs was event data
with 64 energy channels resolution over the full bandpass and a time resolution
of 16~$\mu s$. No detector identification bits were telemetred. At the time of 
the observations, the on-board gain correction algorithm was 
temporarily turned off (Jahoda et~al. 1996). The gains of the 5 PCA detectors 
are slightly different. Whenever we quote intensities in a certain energy 
range, we refer to the energy range consistent with the channel setting of 
detector ``PCU0''. We note that the gains of the other detectors are deviating 
by less than 0.1 keV at the low end to at most 2 keV at the high end of the 
range. 

There are a number of known contributors to the X-ray background in our 
measurements: the cosmic diffuse background, the diffuse galactic ridge
emission and the supernova remnant W49B. The latter two have not been
accurately imaged yet in our bandpass. This forces us to employ a simplified
background subtraction procedure which has been detailed in ISB97: the X-ray 
background is defined as the residual emission found during dip times when 
there is no apparent pulsed emission. Using this background implies 
two assumptions: if there is residual emission from 4U~1907+09 we disregard that and 
assume it is constant, and there is no other variable source in the background.
The last assumption seems reasonable given the mentioned contributors, 
certainly on the time scale of the observations. With regards to residual
emission from 4U~1907+09, we can only note that estimates of the different 
contributions are prone to systematic errors but suggest that residual
emission is at a level less than 2\% of the average 4U~1907+09 intensity in the 
2 to 15~keV band (ISB97).


\section{The X-ray light curve}
\label{lightcurve}

\begin{figure}[t]
  \begin{center}
    \leavevmode
\epsfxsize=13cm
\epsfbox{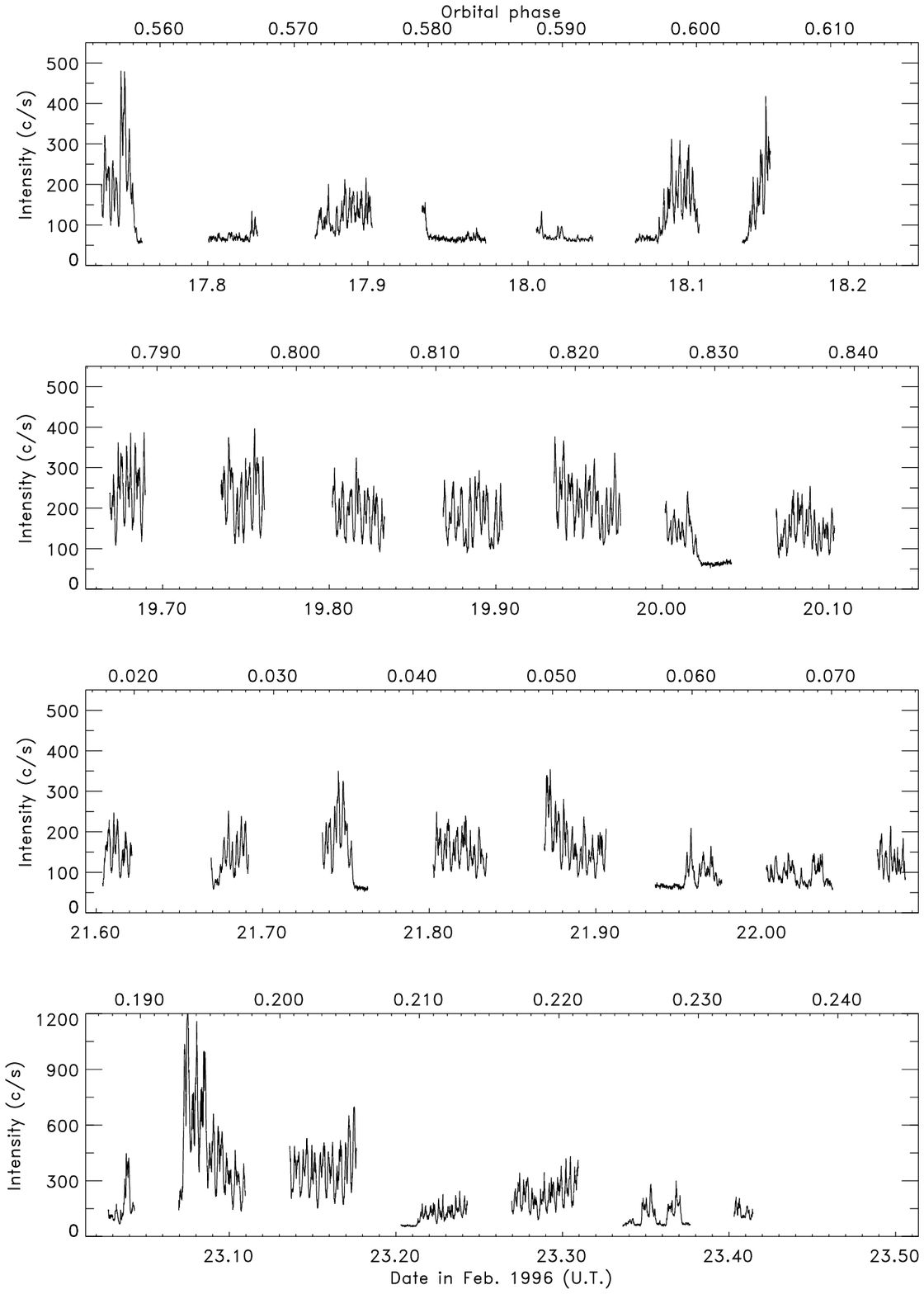}
  \caption{
Time histories of the raw RXTE-PCA 2 to 15 keV intensity of 4U~1907+09,
the bin time is 32~s. The orbital phases are with respect to the 
epoch of the maximum distance between the pulsar and the solar barycenter
\label{fig01}
}
  \end{center}
\end{figure}

\placefigure{fig01}

Figure~\ref{fig01} shows the time history of the raw countrate of the RXTE-PCA
observations on 4U~1907+09. Apart from the dips, the pulsar signal with a period of 
about 440~s is obvious. The pulse profile is very variable. It can change from 
pulse to pulse or even within one pulse. The time history shows a flare at 
Feb.~23.07 when the net 2 to 15 keV intensity rises up by an order of
magnitude to about 0.1~Crab units. Given the newly
determined orbital period (\S~\ref{sectionaltan}) this flare occurs 
543.99$\pm0.04$ orbital periods after the secondary and 544.44$\pm0.04$
orbital periods after the primary flare as observed with Tenma in August and
September of 1983 (M84) which are the most recent timing reports of flares. 
The identification of the flare on Feb.~23.07 with the secondary flare seems
unambiguous and, therefore, we confirm the recurrence of a phase-locked
secondary flare. It is interesting to note that the orbital phase of
the flare (see Table~\ref{tab1}) places it near to the time of apastron 
of the binary. The intensity of the flare is, in 2 to 30 keV,
about twice as high as the one measured with Tenma (M84). This is not 
unexpected. The secondary flare has been seen to vary from insignificant
to as bright as the primary flare (MR80). We note that the observed flare
may actually be only part of a more extended secondary flare which could
have been missed because of the sparsity of the RXTE coverage. The Tenma
observation of the secondary flare suggests the flare may last 0.3 days,
the associated uncertainty in the epoch has been included in the 
above-mentioned uncertainty of the number of orbital periods since the 
Tenma-observed flares.


\section{Timing analysis of the pulse signal}
\label{sectionaltan}

\begin{figure}[t]
  \begin{center}
    \leavevmode
\epsfxsize=13cm
\epsfbox{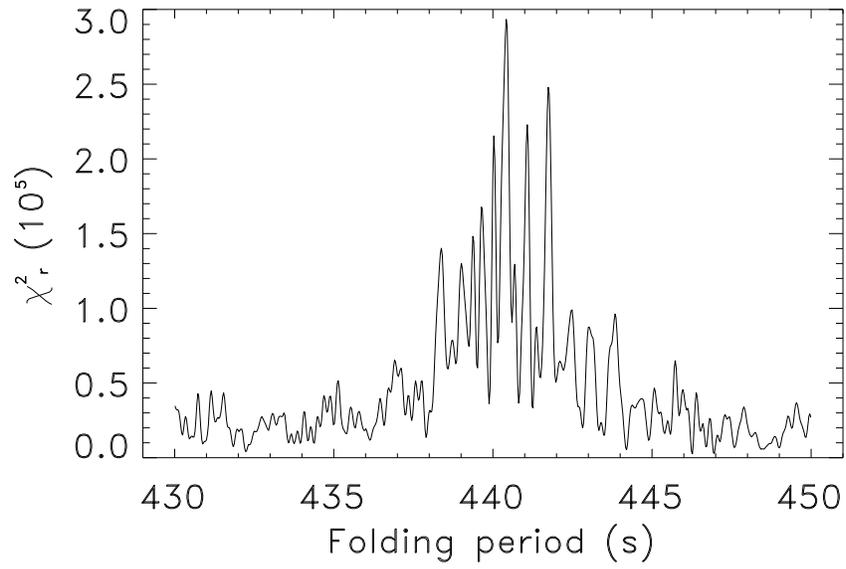}
  \caption{Periodogram of the 2 to 15 keV time history data after 
correction to solar system barycenter but before correction to binary
system barycenter
\label{fig02}}
  \end{center}
\end{figure}

An initial estimate of the pulse period was obtained by folding
the 2 to 15~keV time history on a number of statistically
independent trial periods (Leahy et~al. 1983) in the range 430 to
450~s. Only data outside the intensity dips were used and photon
arrival times were corrected into those for the solar system barycenter.
The highest $\chi^2$ value was found for a period of 440.4~s
(figure~\ref{fig02}). 

\placefigure{fig02}

In order to accurately determine the pulse period as well as the binary orbit 
a set of 19 pulse arrival times was generated, one pulse arrival time for each 
XTE orbit when 4U~1907+09 is not in a dip. This was done by
folding the time history data into one average pulse for each RXTE orbit, 
folding {\em all} time history data into one master pulse, and cross 
correlating the master pulse with each average pulse to find the pulse 
arrival time for each average pulse. Average profiles rather than individual 
pulses were used to minimize 
systematic effects due to the sometimes strong changes in the profile from
pulse to pulse which are supposedly due to variability in the accretion rate
(ISB97). 

As an alternative to control the pulse profile variability we have used the
method of pulse wave filtering as proposed by Deeter \& Boynton (1985,
see also Boynton et~al. 1986).
In this method pulse profiles are expressed in terms of harmonic series and 
cross correlated with the average pulse profile ('master pulse'). The
maximum value of the cross correlation is analytically well-defined and 
does not depend on the phase binning of the pulses. The short term sharp
fluctuations of pulses are naturally filtered by a cut off of higher
harmonics. The pulse arrival times obtained through this method gave 
statistically the same results as those obtained through the above-mentioned 
method.

In order to increase the accuracy of the orbital solution (in particular
the orbital period), the pulse arrival times were combined with the pulse 
delay-time data from Tenma observations as published graphically by M84.

We modeled the data with an eccentric orbit with parameters that have not 
changed since the Tenma observations in September 1983 and with a pulse 
period which is constant throughout our RXTE observations. 
We determined the parameters by testing the model against the data
using Pearson's $\chi^2$ test on a grid of parameter values sufficiently 
sampled to detect significant changes in $\chi^2$ and with a range to
sufficiently enclose the allowed parameter values. The parameters are
the pulse period $P_{\rm pulse}$ at the time of the RXTE observations, 
the orbital period $P_{\rm orb}$, the epoch 
of mean longitude 90 degrees $T_{\pi/2}$ (i.e., one quarter of the orbital 
period after the time of ascending node when the neutron star crosses the sky 
tangent plane through the barycenter moving away from the observer), the 
longitude of periastron 
$\omega$ (i.e., with respect to the ascending node), the eccentricity $e$, 
and the length of the projected semi-major axis $a_{\rm x}$~sin~$i$/c. The grid
ranges and sample frequencies were found iteratively, going from a rough to
a sufficiently fine grid. We choose this grid-search method to be able to 
determine the uncertainty of the solution with any arbitrary confidence level. 
The $\chi^2$ values were calculated in the pulse delay-time domain. On the one
hand this enables the inclusion of the Tenma data that have not been published
in the pulse arrival-time domain, on the other hand this precludes the
determination of $P_{\rm pulse}$ during the Tenma observations.  We fixed
the error in the pulse delay times at 8~s. This value was suggested by
the uncertainty of the harmonics in the pulse wave filtering analysis of
the RXTE data and was the value M84 applied for the Tenma data.

\begin{figure}[t]
  \begin{center}
    \leavevmode
\epsfxsize=13cm
\epsfbox{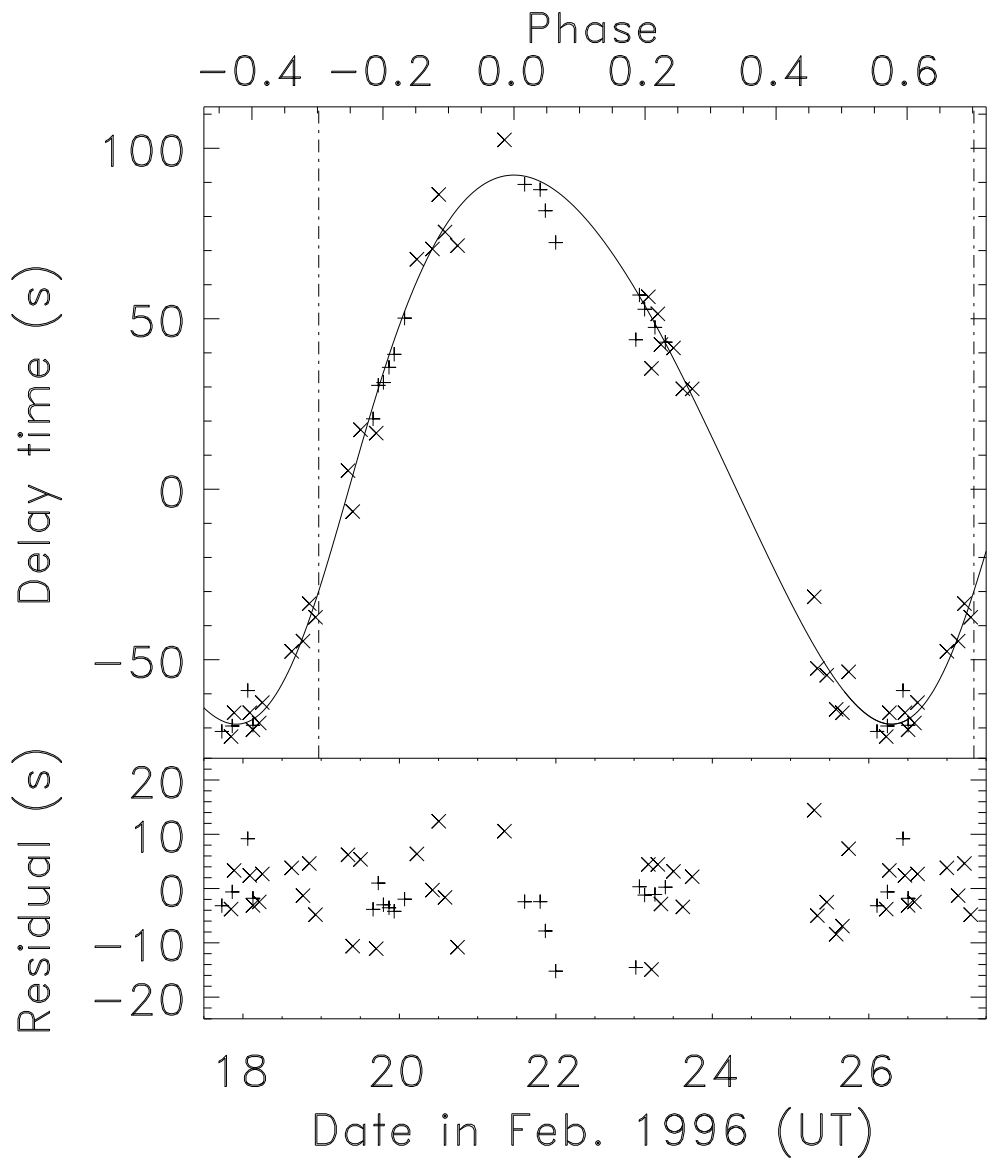}
  \caption{The delay time versus time for the RXTE measurements (+ symbols) and
for the Tenma measurements folded into the RXTE time of observation 
($\times$ symbols). The phases indicated on top are with respect
to the time of maximum distance between the pulsar and the solar barycenter. 
The solid line shows the model for the eccentric orbit
specified in Table~\ref{tab2}. The vertical dashed lines refer to the 
inferred times of periastron. All data points are repeated modulo 
$P_{\rm orb}$. The rms of the residuals is 6.6~s
\label{fig03}}
  \end{center}
\end{figure}

The results of modeling 
the timing data are presented in Table~\ref{tab2}. The quoted error intervals
are the projections of the 68\% confidence
level region onto each of the 6 parameter axes. Since some parameters are 
very dependent, the actual solution is confined within a much smaller space
than that of the 6-dimensional cube dictated by the quoted errors. The errors
are, therefore, conservative. For the purpose of easy comparison with orbital
solution by others, we have also included the single-parameter 1$\sigma$ errors
in Table~\ref{tab2}. Figure~\ref{fig03} presents a plot of the orbital 
model versus the data in terms of delay times with respect to the binary 
barycenter.

\placefigure{fig03}

The orbital solution provides a means to correct the pulse period as determined
with Ginga in 1990 (Mihara 1995) for the orbital motion of the neutron star in 
the binary
system. The Ginga observations were performed for a duration of 14 hours (this 
is 0.07 in phase) at a mean time of MJD~48156.60. This is 0.10 orbital periods
after the epoch of ascending node where the orbital Doppler shift is
+0.28~s per pulse period with an uncertainty of about 0.01~s per pulse period. 
Therefore, the corrected pulse period is 439.19~s.

We find a pulse period at the time of the RXTE observations of 
$440.341^{+0.012}_{-0.017}$~s. This completes a total of four measurements
of the pulse period and 3 measurements of the pulse period derivative since 
1983. These values, tabulated in Table~\ref{tab3}, imply 
an interesting finding: the pulsar appears to be spinning down with a close to
constant derivative. The three values are within 8\% of a mean value
of $\dot{P}_{\rm pulse}=+0.225$~s~yr$^{-1}$.

As mentioned, we assumed that over the course of 12 years there is no
noticeable change in $P_{\rm orb}$, $\omega$ and the orbit inclination angle 
$i$. For $P_{\rm orb}$, we can confirm this with reasonable accuracy because
the value we find is, within error margins, equal to what was found by CP87
between the Tenma and EXOSAT measurements (8.3745$\pm0.0042$ days). 
For $\omega$, the uncertainty,
typically 20 degrees, is simply too large to be at all able to measure 
likely values for the change (for a detailed analysis of
apsidal advance in the similar system Vela X-1, see Deeter et~al.
1987). The same argument holds for the sensitivity towards measuring changes 
in~$i$.

\begin{figure}[t]
  \begin{center}
    \leavevmode
\epsfxsize=8cm
\epsfbox{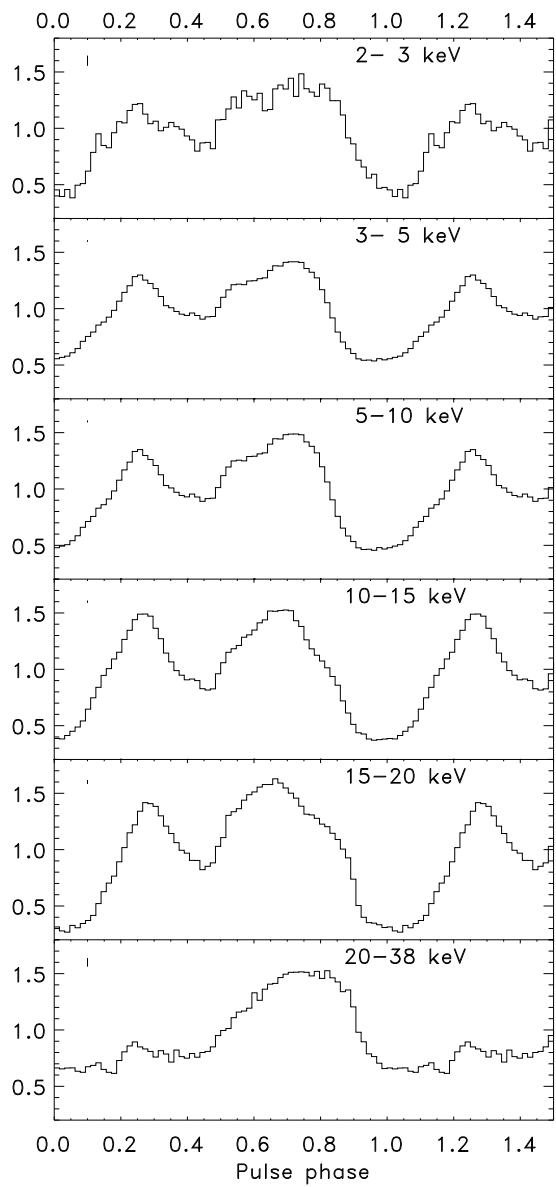}
  \caption{The background-subtracted lightcurve folded with the pulse period for 6 
bandpasses between 2 and
38~keV. The bandpasses are indicated in each panel as well as the 
statistical 1$\sigma$ error in the upper left corner (from top to bottom
these errors are equal to 0.045, 0.007, 0.009, 0.010,
0.018 and 0.038). The phase offset is
arbitrary. The unit of intensity is the average intensity per bandpass. Only
data for 'quiet' periods is used when the source is neither dipping nor
flaring above 500 c~s$^{-1}$. The net exposure time for these lightcurves
is 29.6~ks out of a total of 79~ks
\label{fig04}}
  \end{center}
\end{figure}

\placefigure{fig04}

\begin{figure}[t]
  \begin{center}
    \leavevmode
\epsfxsize=13cm
\epsfbox{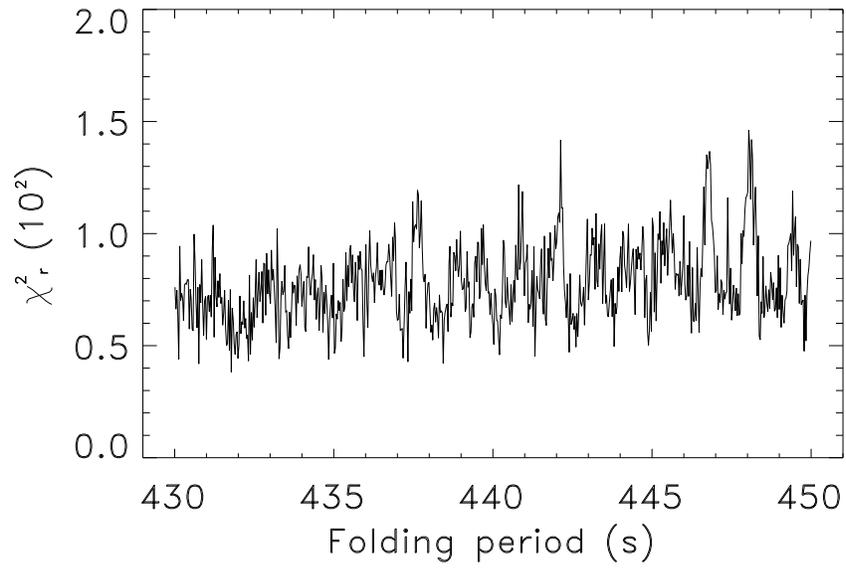}
  \caption{Periodogram of the $>40$ keV time history data after 
correction to solar barycenter but before correction to binary
barycenter
\label{fig05}}
  \end{center}
\end{figure}

\placefigure{fig05}

The photon arrival times were folded with the pulse period after correction
for the earth's motion around the sun, RXTE's motion around the 
earth and the binary motion of the pulsar. The resultant pulse profile
is presented in figure~\ref{fig04} for 6 photon energy bands
up to 38 keV. Above this energy no pulsed emission was detected, as revealed
by a Fourier analysis (figure~\ref{fig05}). The pulse profiles are 
expressed in units of 
average 4U~1907+09 intensity per band. The background in each band was determined
as discussed in ISB97 (see \S\ref{observations}). 
The pulse profile consists of two peaks with a deep and a shallow minimum in
between that are close to 0.5 in pulse phase apart. The pulse profile looks 
similar to that observed more than a decade before with Tenma and EXOSAT in
similar energy bands (M84 and CP87). It is rather insensitive 
to energy between 2 and 20~keV although subtle dependencies are noticeable, 
particularly in the second peak. 
There is a dramatic change of the pulse profile at around 20~keV. Basically, 
the first pulse disappears above that energy and the shape of the other one 
changes. The modulation depth is roughly the same in all bands.
The energy dependency of the pulse profile appears similar to that observed 
from Cen~X-3 with Ginga (Nagase et~al. 1992). 


\section{Transient 18~s oscillations}
\label{sectiontod}

\begin{figure}[t]
  \begin{center}
    \leavevmode
\epsfxsize=13cm
\epsfbox{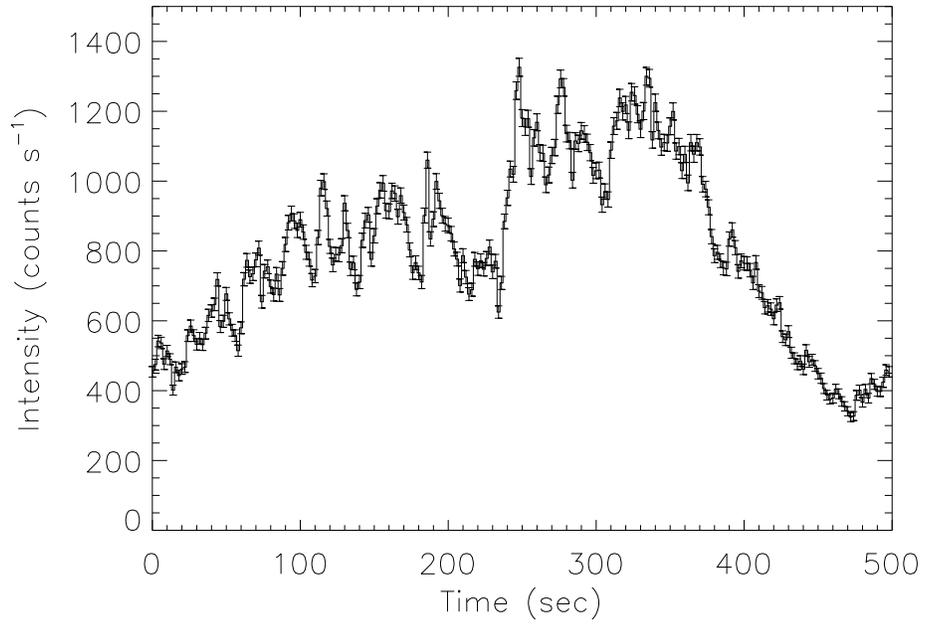}
  \caption{An expanded view of the observed X-ray time profile during the secondary
flare in the full PCA bandpass. Time is measured from Feb. 23, 1996,
01:56:33 UTC
\label{fig06}}
  \end{center}
\end{figure}

\placefigure{fig06}

\begin{figure}[t]
  \begin{center}
    \leavevmode
\epsfxsize=13cm
\epsfbox{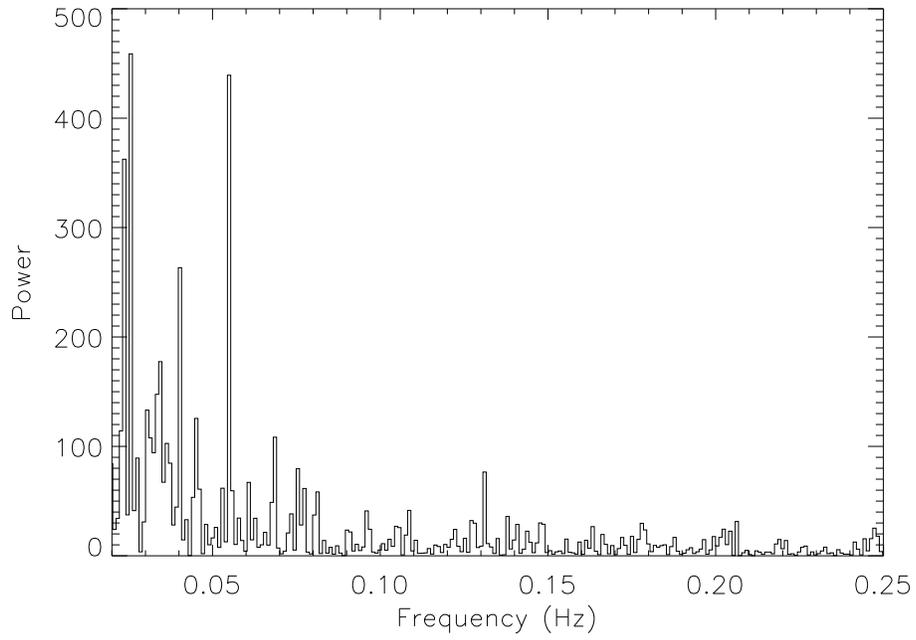}
  \caption{The power spectrum of a 1024~s piece of the lightcurve starting 300~s
before the start time indicated in figure~\ref{fig06}
\label{fig07}}
  \end{center}
\end{figure}

\placefigure{fig07}

\begin{figure}[t]
  \begin{center}
    \leavevmode
\epsfxsize=13cm
\epsfbox{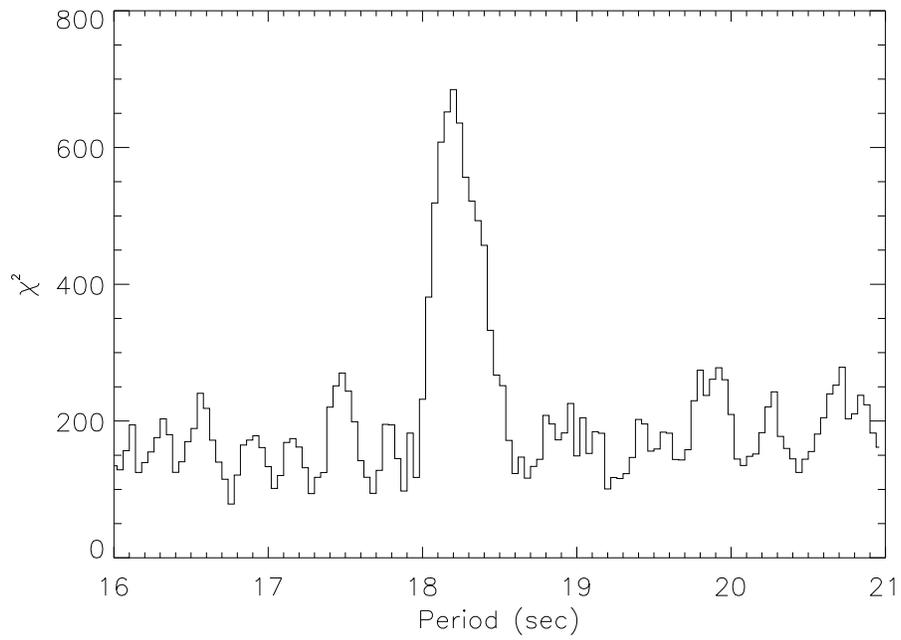}
  \caption{Periodogram near 18~s for the flare data
\label{fig08}}
  \end{center}
\end{figure}

\begin{figure}[t]
  \begin{center}
    \leavevmode
\epsfxsize=13cm
\epsfbox{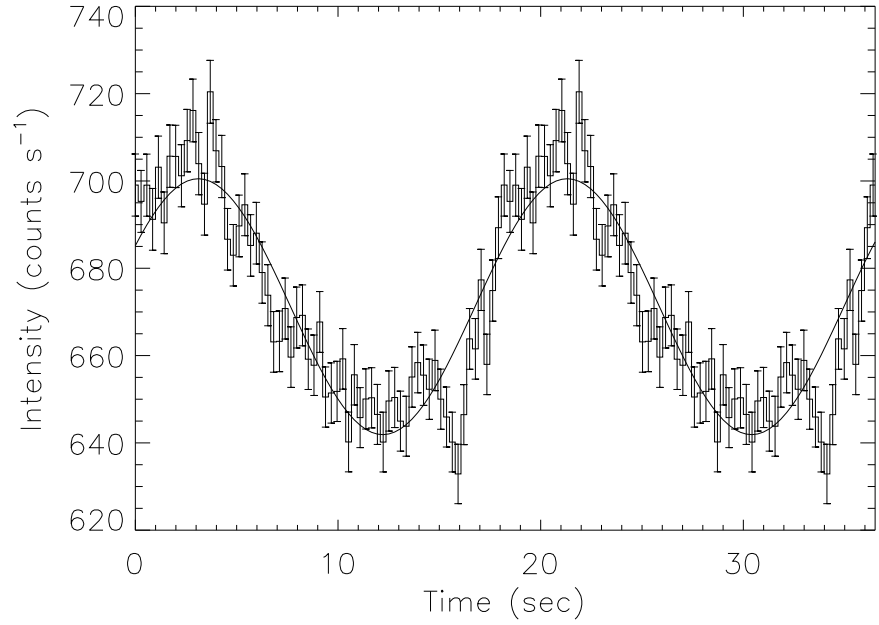}
  \caption{Folded background-subtracted lightcurve, folded with a period of 18.2~s,
and a sinusoidal fit to the data
\label{fig09}}
  \end{center}
\end{figure}

Visual inspection of the X-ray lightcurve on second timescales during the 
flare at Feb. 23.07 suggests that there is a variety of non-Poissonian variability 
present during the flare. For example, figure~\ref{fig06} shows the lightcurve
of a 500 second interval near the peak of the flare. Fluctuations in the 
countrate as large as several hundred counts~s$^{-1}$ are obvious in this 
lightcurve. To further quantify this 
short timescale variability (i.e., shorter than a pulse period) we computed
Fourier power spectra for the time intervals encompassing the flare.
The power spectrum of a 1024 second interval beginning 300 seconds prior to
that shown in figure~\ref{fig06} is displayed in figure~\ref{fig07}. There are a number of
conspicuous peaks in this power spectrum in the range from 0.02 to 0.06~Hz.
There is also clearly a broadband noise component increasing toward lower
frequencies, as the mean power is significantly above 2 (the mean expected
for purely poisson fluctuations) extending above 0.2~Hz. One of the most
prominent peaks is at 0.055~Hz, or a period of 18.2~s. These 18 s oscillations
can actually be seen with the eye in figure~\ref{fig06}. To investigate further we
performed an epoch folding period search in the vicinity of 18~s. The results are
shown in figure~\ref{fig08}, and there is an obvious peak centered at 18.2 s.  
To determine the average amplitude of the oscillations, we folded the 1024 s of
data on the best-fit period of 18.2 seconds . We then fit a model including a
constant countrate plus a sinusoid to the folded data. Figure~\ref{fig09} shows the
resulting background subtracted lightcurve and the best-fit sinusoidal model. 
The $\chi^2$ per degree of freedom is formally a bit high, 1.8, but for estimating 
the amplitude of the oscillation the sinusoidal fit is sufficient.
From this analysis we obtain an average oscillation amplitude of $4.4 \pm
0.3\%$, where the amplitude is defined as the ratio of the sinusoidal to
constant countrate components. We also investigated the dependence on photon 
energy of both the pulsation amplitude and pulse profile, but found no
significant energy dependence.

\placefigure{fig08}

\placefigure{fig09}

\begin{figure}[t]
  \begin{center}
    \leavevmode
\epsfxsize=13cm
\epsfbox{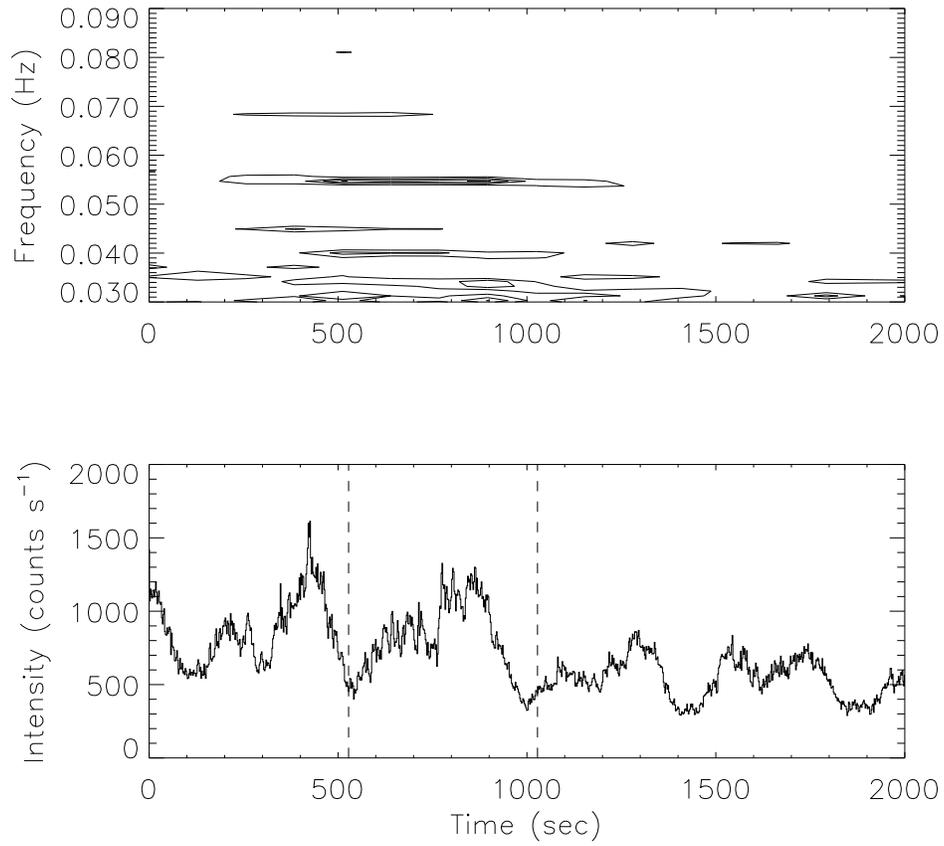}
  \caption{The top panel shows a contour map of the dynamic power spectrum. The bottom
panel shows the accompanying lightcurve. The vertical dashed lines indicate
the time frame of figure~\ref{fig06}
\label{fig10}}
  \end{center}
\end{figure}

The 18~s oscillations are not persistent, rather, they are confined to an 
approximately 1000 s interval centered near the peak of the flare. To 
investigate the transient nature of the 18 s oscillation we computed a
dynamic power spectrum by calculating power spectra from 500 second intervals
with a new interval beginning every 50 seconds. The resulting power spectra are
not independent, since the data segments overlap, however, this method
identifies the range of times for which the 18~s oscillations are present. 
The resulting dynamic spectrum is shown in the top panel of figure~\ref{fig10}.
The 0.055 Hz oscillations are clearly present from about
200-1200 seconds, and this is the only epoch in which we detected such 
oscillations. During the time that the oscillations are present there is
no strong evidence for significant frequency drift, this gives a lower limit
to the $Q$-value for the oscillation of $Q = (1000 {\rm s} / 18.2 {\rm s}) = 
55.6$ ($Q$ is
the so-called quality factor which measures the number of oscillations that will
pass before a substantial fraction of the energy of the oscillator is dissipated, 
$Q$ is inversely proportional to the bandwidth or spectral purity of the 
oscillation).

\placefigure{fig10}

\begin{figure}[t]
  \begin{center}
    \leavevmode
\epsfxsize=13cm
\epsfbox{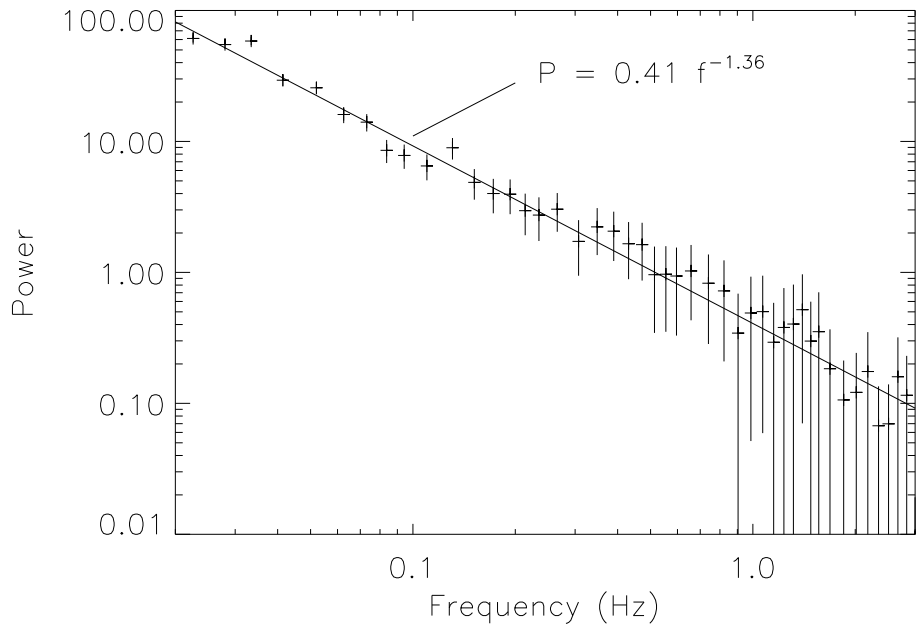}
  \caption{The power spectrum of the broadband noise
\label{fig11}}
  \end{center}
\end{figure}

To investigate the nature of the broadband noise we computed an average FFT
power spectrum from two successive orbits during the flare, for a total of 6250
seconds of data. The resulting average power spectrum is shown in figure~\ref{fig11}.
We fit a power law model $P = K \nu^{-\alpha}$ to 51 frequency bins in the 0.02
to 4 Hz frequency range. We did not fit below 0.02 Hz because there the power
spectrum is dominated by the pulsed signal and its harmonics. The power law
model with $K = 0.41 \pm 0.07$ and $\alpha = 1.36 \pm 0.06$ provides an
acceptable fit, with a $\chi^2 / {\rm d.o.f.} = 39/49 \approx 0.8$. The integrated power
from 0.02 to 4 Hz corresponds to an amplitude (rms) of about 8.2\%.

\placefigure{fig11}

We note that we have searched for pulsations at higher frequencies and failed
to find any. No pulsations are present like were reported by Sadeh \& Livio
(1982) at a period of around 15.3~ms. During the flare we find an upper limit
for the amplitude of 0.5\% as compared to the amplitude of 12\% at which
pulsations were seen by Sadeh \& Livio.


\section{Discussion}
\label{discussion}

Quasiperiodic oscillations (QPO) with frequencies in the 10 - 200 mHz range
have been observed from 7 other X-ray pulsars. Among these, A0535+26 (Finger, 
Wilson, \& Harmon 1996), X1626-67 (Shinoda et~al. 1990), Cen X-3 
(Takeshima et~al. 1991), V0332+53 (Takeshima et~al. 1994), 
X0115+63 (Soong \& Swank 1989), and SMC X-1 (Angelini 1989) all have QPO
frequencies in the 0.062 to 0.1 Hz regime, similar to the 0.055 Hz 
transient oscillation from 4U~1907+09 described above. The $\approx 4\%$
amplitude (rms) of the 18 s oscillation is also similar to the amplitudes of 
QPO from other X-ray pulsars (see Takeshima et~al. 1994; and Angelini,
Stella \& Parmar 1989), however, QPO detected in these sources typically are
broader, with $Q$-values of order a few compared to about 50 for the 18 s
oscillation in 4U~1907+09.

To date, almost all models for QPO in X-ray pulsars postulate the existence of
an accretion disk as the site of QPO production.  In the case of QPO from A0535+26 
this is supported by the observed correlation of the QPO frequency
and spin-up rate (Finger, Wilson, \& Harmon 1996), strongly suggesting the 
transfer of angular momentum from accretion disk to neutron star. Thus, one
possibility is that the 18 s oscillations reveal the presence of a transient 
accretion disk during the flare. 

Assuming that the 18 s oscillation in 4U~1907+09 is related to orbital
motion, either via a beat frequency model (BFM, Lamb et~al. 1985; Alpar 
\& Shaham 1985), or a Kepler frequency model (KFM, van der Klis et~al.
1987), then the inferred radius of material in a putative disk is $R_{\rm d} =
(GM/4\pi^2\nu_{QPO}^2)^{1/3} \approx 1.2 \times 10^{4}$ km, for a 
$1.4$~$M_{\sun}$
neutron star. This is nearly an order of magnitude smaller than the co-rotation
radius $R_{\rm c} = 9.6 \times 10^{4}$ km for this object. We can rougly 
estimate the size of the neutron star magnetosphere from the expression for 
the Alfv\'{e}n radius assuming spherical accretion (e.g., Ghosh \& Lamb 1991),
\begin{eqnarray}
R_{\rm m} & = & 2.4\times 10^3 (M/1.4~M_{\sun})^{1/7} \times\nonumber\\
          &   & (B/2.5 \times 10^{12} G)^{4/7} \times\nonumber\\
	  &   & (R/10^6~{\rm cm})^{10/7} \times\nonumber\\
	  &   & (L_{\rm x}/3.1\times 10^{37} {\rm ergs~s}^{-1})^{-2/7}\hspace{0.3cm}{\rm km}.
\end{eqnarray}
Studies of the optical counterpart, discovered by Schwartz et~al. (1972), 
reveal that the distance to 4U~1907+09 is between 2.4 and 5.9~kpc
(Van Kerkwijk~et~al. 1989). Assuming the distance is 4~kpc, the average 
non-absorbed luminosity of 4U~1907+09 during non-dip and non-flare periods and within 
the studied energy range is $1\times10^{36}$~erg~s$^{-1}$ (ISB97). During the 
flare the luminosity increases up to $4\times10^{36}$~erg~s$^{-1}$. 
For a flare X-ray luminosity of $4 \times 10^{36}$ ergs/s, a magnetic field 
$B = 2.5 \times 10^{12}$~G and a stellar mass and radius of $1.4$~$M_{\sun}$ and 
10 km respectively, $R_{\rm m} \approx 4300$ km. Although this number is rather
uncertain it suggests that
the disk can certainly penetrate close enough to the star to account for the
18 s oscillations as an orbital phenomenon associated with an accretion disk.
Although the accretion disk interpretation seems plausible, further
observations, particularly in the vicinity of the flare will be required to 
confirm this interpretation.

Long-term measurements of the behavior of the spin period for a selection of 15 
accreting X-ray pulsars (Bildsten et~al., 1997) show that three known pulsars 
appear to exhibit systematic spin-down evolution for at least 5 years: GX1+4 
($P/\dot{P}\approx$90~yrs), 4U~1626-67 ($P/\dot{P}\approx5\times10^3$~yrs)
and Vela~X-1 ($P/\dot{P}\approx6\times10^3$~yrs). Generally, this behavior
is attributed to low accretion rates that bring the magnetospheric radius 
$R_{\rm m}$ close to the co-rotation radius $R_{\rm c}$ (Ghosh \& Lamb 1979a, 
1979b; Wang 1987, 1995; Anzer \& B\"{o}rner
1980; Lovelace, Romanova \& Bisnovatyi-Kogan 1995). However, for 4U~1907+09 
this is an unlikely scenario because the magnetospheric radius is hard to bring
out to such a large co-rotation radius unless either the magnetic field is of 
unlikely order 10$^{14}$~G or the distance is of order 
0.5~kpc which is at least 5 times lower than studies of the optical counterpart
suggest.

The putative disk supports the notion that the magnetospheric radius
is substantially smaller than the co-rotation radius. Moreover, a disk
can provide an alternative explanation to the observed spin down if
this transient disk is rotating in an opposite sense to the pulsar spin. 
Such a disk can provide the 
necessary torque to explain the spin down rate. If all accreted mass supplies 
its angular momentum at a radius $R_{\rm d}$ the expected torque on the neutron
star is $N_{\rm char}=\eta\dot{M}(GMR_{\rm d})^{1/2}$ where $\eta<1$
is the duty cycle of the transient disk. If we assume that all the 
liberated potential energy of the accreted mass during the flare is 
transformed into radiation and the flare luminosity is 
$3.2\times10^{35}D_{\rm kpc}^2$~erg~s$^{-1}$, then 
$N_{\rm char}=2.5\times10^{22}\eta D_{\rm kpc}^2R_{\rm d}^{1/2}$. For
$R_{\rm d}=1.2\times10^4$~km this becomes 
$N_{\rm char}=8.6\times10^{32}\eta D_{\rm kpc}^2$~g~cm$^2$.
The observed absolute value of the torque is
$N_{\rm obs} = 2\pi~I|\dot{\nu}|$ where I is the moment of inertia and 
$\dot{\nu}=-3.7\times10^{-14}$~Hz~s$^{-1}$ (equivalent to 
$\dot{P}=+0.225$~s~yr$^{-1}$). If $I=10^{45}$~g~cm$^2$ (generic value for
a neutron star of radius 10~km and mass 1.4~M$_{\sun}$), 
$N_{\rm obs}=2.3\times10^{32}$~g~cm$^2$. If $N_{\rm obs}=N_{\rm char}$ and 
$2.4<D_{\rm kpc}<5.9$, then $0.008<\eta<0.05$. Therefore, the putative
retrograde transient accretion disk can provide the torque to spin down
the pulsar if it lasts on average on the order of a few percent of the time.
We observed a duty cycle of about 1000~s out of an orbital period of 8.4~d
which is an order of magnitude too small. However, the coverage of our 
observations is limited and also the flare might, averaged over many orbits,
have a larger duty cycle. This is confirmed by Tenma observations which 
suggest a duration of 0.3~d for one flare (M84). We conclude that the putative 
retrograde accretion disk could possibly supply enough negative torque to 
spin down the pulsar.

The suggestion of a retrograde accretion disk being responsible for an 
extensive spin down period has recently been revisited in the case of the 
disk-fed AXP GX1+4 by Chakrabarti et~al. (1997) in order to have an 
elegant explanation for a positive correlation between luminosity and negative
torque. Our need for a retrograde accretion disk in the wind-fed 4U~1907+09 is 
motivated by a co-rotation radius which is clearly much larger than the
magnetospheric radius.

4U~1907+09 is in several ways similar to Vela~X-1: the spin periods are both of
order several hundred seconds (440~s for 4U~1907+09 and 283~s for Vela~X-1), the 
orbital periods are of order 8 days (8.38~d for 4U~1907+09 and 8.96~d for Vela X-1), 
and both have recently been found to show long periods of systematic spin down
(e.g., Bildsten et~al. 1997 for Vela~X-1). Previously, the occasional spin-down 
periods in Vela~X-1 have been attributed to the presence of a disk with an inner
edge close to the co-rotation radius (Nagase 1989), with the note that this 
implies relatively large magnetic fields (surface fields of order 10$^{13}$~G).
We suggest that, since this reasoning goes awry in the case of 4U~1907+09,
this conclusion may be invalid for Vela~X-1 as well. An important difference
between both systems is that Vela~X-1 shows much more spectral variability
and pronounced iron lines. This may be due to a difference in angle of sight
of the binary orbit, a difference in the spatial distribution of
the wind from the companion star, and/or a somewhat different eccentricity.

There is, as compared with some other AXPs, moderate spectral variability within 
the pulse profile. Basically the pulse profile shows two peaks below and one peak 
above $\sim$20~keV. This is somewhat similar to what has been observed from the 
much more luminous disk-fed AXP Cen~X-3 and may be associated with two magnetic 
poles which have either different physical circumstances or are viewed on at 
different angles by the observer. A modeling of the pulse profile, similar to 
studies performed by Bulik et~al. (1995) and beyond the scope of this paper, may
provide more definite conclusions about this.

Since its discovery by MR80, the occurrence of two flares per orbit has
been difficult to reconcile with the identification of the companion as
a supergiant high-mass star (e.g., Van Kerkwijk et~al. 1989, ISB97). It has
always been thought that these stars do not have a circumstellar disk
like is presumed for Be stars. However, recently evidence is mounting that
in hot supergiants in the upper part of the H-R diagram axial symmetry may
play an important role (e.g., Zickgraf et~al. 1996 and references therein).
Also, recently another wind-fed AXP with a 
supergiant companion has been shown to consistently display two flares per 
orbit: GX~301-2 (Pravdo et~al. 1995; Chichkov et~al. 1995; Koh et~al. 1997). This again
supports the notion that the wind from the companion is not isotropic but
that the mass flux is enhanced along the equatorial plane. The two flares
then may be caused by the neutron star traversing this plane twice per orbit
in a sufficiently inclined orbit. 
It is interesting to note in this context 
that GX~301-2 (Sato et~al. 1986) and 4U~1907+09 have the highest eccentricity of the group of 
supergiant high-mass XRBs, perhaps these systems are not yet in a tidal 
equilibrium. Perhaps even the spin orientation of the neutron star differs
by more than 90$^{\rm o}$ from the orbital motion orientation. However,
at the moment this is mere speculation.


\section{Conclusion}
\label{conclusion}

We have determined the pulse period of 4U~1907+09 at a recent epoch and find 
that the pulsar has spun down on average by 0.225~s~yr$^{-1}$ since the
pulsar's discovery in 1983. Three measurements of the spin down rates
during intermediate time interval spanning between 1 year and 6 years 
are consistent within 8\% of this value.
This suggests a remarkable monotonous spin down trend during about 12 years.
Furthermore, we find the occurrence of 18~s oscillations for 10$^3$~s
of a flare with a high Q~value hinting at an accretion disk.
The indication that the magnetospheric radius is much smaller than the 
co-rotation radius, the observed long and constant spin down trend and the 
occurrence 
of transient oscillations suggest that possibly a recurrent transient
accretion disk counter rotates the neutron star and, through transfer of 
angular momentum, slows the pulsar down. This suggestion needs to be
confirmed by at least detailed multiple observations of flares which are 
almost certainly recurring every orbital period.


\acknowledgements
We thank the members of the RXTE Guest Observer Facility for their help in
the data reduction and the anonymous referee for useful suggestions. 
J.Z. and A.B. acknowledge the support under the 
research associateship program of the U.S. National Research Council during
their stay at NASA's Goddard Space Flight Center when part of the
research presented in this paper was performed. T.S. acknowledges 
the support of the HEAP program of the Universities Space Research Association
at NASA's Goddard Space Flight Center.


\newpage

\begin{table}
\caption[]{RXTE observation log of 4U~1907+09}
\label{tab1}
\centering
\begin{tabular}{r|rrrr}
\hline
Observation run & 1 & 2 & 3 & 4 \\
Dates (Feb. 1996, U.T.) & 17.73-18.15 & 19.67-20.10 & 21.60-22.08 & 23.02-23.42 \\
Orbital phases\tablenotemark{a} & 0.55-0.60 & 0.78-0.84 & 0.01-0.07 & 0.18-0.23\\ 
Orbital phases\tablenotemark{b} & 0.85-0.90 & 0.08-0.14 & 0.31-0.37 & 0.48-0.53 \\
Exposure time (s)   & 19442 & 19911 & 20261 & 19775 \\
Time span (s)	    & 36093 & 37601 & 41658 & 33457 \\
\hline
\end{tabular}

\noindent
\tablenotetext{a}{Based on epoch of maximum distance between neutron star and earth}
\tablenotetext{b}{Based on epoch of periastron}
\end{table}

\begin{table}
\caption[]{The binary orbit and pulse period of 4U1907+09 from 1983 Tenma and
1996 RXTE-PCA measurements}
\label{tab2}
\centering
\begin{tabular}{llrrr}
\hline
Parameter & Symbol & Value & 68\%           & Single \\ 
          &        &       & confidence     & parameter \\
          &        &       & region         & 1$\sigma$ error \\
\hline
Orbital period  & $P_{\rm orb}$ & 8.3753 days & $^{+0.0003}_{-0.0002}$ & 0.0001\\
Eccentricity    & $e$           & 0.28        & $^{+0.10}_{-0.14}$ & 0.04 \\
Orbital epoch   & $T_{\pi/2}$   & MJD 50134.76& $^{+0.16}_{-0.20}$ & 0.06 \\
Longitude of periastron & $\omega$ & 330 degrees & $^{+20}_{-20}$ & 7 \\
Projected semi-major axis length & $a_{\rm x}$ sin~$i/c$ & 83 lt-s & $^{+4}_{-4}$ & 2 \\
Pulse period  & $P_{\rm pulse}$ & 440.341 s & $^{+0.012}_{-0.017}$ & 0.006\\
\hline
\end{tabular}
\end{table}

\begin{table}
\caption[]{History of $P_{\rm pulse}$ measurements for 4U~1907+09}
\label{tab3}
\centering
\begin{tabular}{llllll}
\hline
\multicolumn{2}{c}{Mean time}& Satellite & Reference & P$_{\rm pulse}$ & Derivative\tablenotemark{a} \\
Date            & (MJD)      &           &           & (s)          & (s~yr$^{-1}$)\\
\hline
Aug./Sept. 1983	& 45576      & Tenma	 & M84	     & 437.483$\pm$0.004& ---            \\
May/June 1984	& 45850      & EXOSAT	 & CP87	     & 437.649$\pm$0.019& 0.22$\pm$0.03  \\
Sep. 1990	& 48156.6    & Ginga	 &Mihara 1995& 439.19$\pm$0.02\tablenotemark{b}  & 0.244$\pm$0.005\\
Feb. 1996	& 50134	     & RXTE	 &this paper & 440.341$\pm$0.014& 0.212$\pm$0.004\\
\hline
\end{tabular}

\noindent
\tablenotetext{a}{The pulse period derivative is calculated from the difference in $P_{\rm pulse}$
	with respect to the previous observation in this table.}
\tablenotetext{b}{This value was corrected for delays from binary motion in the present work. The 
	uncertainty is an estimate.}
\end{table}


\begin{references}{}
\reference{Alpar}Alpar, M. A., \& Shaham, J. 1985, Nature, 316, 239
\reference{Angelini89a}Angelini, L. 1989, in Proc. 23d ESLAB Symp., Two Topics in X-ray
	Astronomy, Vol. 1, ed. N. White (Garching: ESA), 81
\reference{Angelini89b}Angelini, L., Stella, L., \& Parmar, A. N. 1989, ApJ, 346, 906
\reference{Anzer80}Anzer, U., B\"{o}rner, G. 1980, \aap, 83, 133
\reference{Bildsten97}Bildsten, L., Chakrabarti, D., Chiu, J., Finger, M.H., 
	Koh, D.T., Nelson, R.W., Prince, T.A., Rubin, B.C., Scott, D.M., 
	Stollberg, M., Vaughan, B.A., Wilson, C.A., Wilson, R.B. 1997, \apjs,
	113, 367
\reference{Boynton86} Boynton, P.E., Deeter, J.E., Lamb, F.K., Zylstra, G.
	1986, \apj, 307, 545
\reference{Bulik}Bulik, T., Riffert, H., Meszaros, P., Makishima, K., 
	Mihara, T., Thomas, B. 1995, \apj, 444, 405
\reference{Chak1}Chakrabarty, D., Bildsten, L., Finger, M.H., Grunsveld, J.M.,
	Koh, D.T., Nelson, R.W., Prince, T.A., Vaughan, B.A., Wilson, R.B.
	1997, \apj, 481, L101
\reference{Chich}Chichkov, M.A., Sunyaev, R.A., Lapshov, I.Y., Lund, N., Brandt, S.,
	Castro-Tirado, A. 1995, PAZh, 21, 491
\reference{Cook} Cook, M.C., Page, C.G. 1987, \mnras, 225, 381
\reference{Deeter85}Deeter, J.E., Boynton, P.E. 1985, Proc. Inuyama workshop 
	on Timing Analysis of X-ray Sources, eds. S. Hayakawa \& F. Nagase
\reference{Deeter87}Deeter, J.E., Boynton, P.E., Lamb, F.K., Zylstra, G. 1987,
	\apj, 314, 634
\reference{Finger}Finger, M. H., Wilson, R. B., \& Harmon, B. A. 1996, ApJ, 459, 288
\reference{Gia} Giacconi, R., Kellogg, E., Gorenstein, P., Gursky, H., 
	Tananbaum, H. 1971, \apjl, 165, L27
\reference{Ghoa}Ghosh, P., Lamb, F.K. 1979a, \apj, 232, 259
\reference{Ghoa}Ghosh, P., Lamb, F.K. 1979b, \apj, 234, 296
\reference{Ghoa}Ghosh, P., Lamb, F.K. 1991, in "Neutron Stars Theory and
	Observation", eds. J. Ventura and D. Pines, NATO ASI Series 43, 363
\reference{Inoue}Inoue, H. 1985, \ssr, 40, 317
\reference{ISB97}In 't Zand, J.J.M., Strohmayer, T.E., Baykal, A. 1997, \apj,
	479, L47
\reference{Jaho} Jahoda, K., J.J. Swank, Giles, A.B., Stark, M.J., 
	Strohmayer, T., Zhang, W. 1996, Proc. SPIE, 2808, 59
\reference{Koh}Koh, D.T., Bildsten, L., Chakrabarti, D., Nelson, R.W.,
	Prince, T.A., Vaughan, B.A., Finger, M.H., Wilson, R.B.,
	Rubin, B.A. 1997, \apj, 479, 933
\reference{Lamb}Lamb, F. K., Shibazaki, N., Alpar, M. A., \& Shaham, J. 1985, 
	Nature, 317, 681
\reference{Leahy} Leahy, D.A. Darbro, W., Elsner, R.F., Weisskopf, M.C.,
	Sutherland, P.G., Kahn, S., Grindlay, J.E. 1983, \apj, 266, 160
\reference{Love}Lovelace, R.V.E., Romanova, M.M., Bisnovatyi-Kogan, G.S. 1995,
	\mnras, 275, 244
\reference{Maki84} Makishima, K., Kawai, N., Koyama, K., Shibazaki, N., 
	Nagase, F., Nakagawa, M. 1984, \pasj, 36, 679
\reference{Maki96}Makishima, K., Mihara, T. 1992, in "Frontiers of X-ray 
	Astronomy", Proceedings of the
	Yamada Conference XXVIII, eds. Y. Tanaka \&
	K. Koyama (Universal Academy Press: Tokyo), 23
\reference{Marsh} Marshall, N., Ricketts, M.J. 1980, \mnras,
	193, 7P
\reference{Mihara} Mihara, T. 1995, Ph.D. thesis, University of Tokyo
\reference{Naga98}Nagase, F. 1989, \pasj, 41, 1
\reference{Nagase} Nagase, F., Corbet, R.H.D., Day, C.S.R., Inoue, H., 
	Takeshima, T., Yoshida, K., Mihara, T. 1992, \apj, 396, 147
\reference{Pravdo}Pravdo, S.H., Day, C.S., Angelini, L., Harmon, B.A.,
	Yoshida, A., Saraswat, P. 1995, \apj, 454, 872
\reference{Sadeh}Sadeh, D., Livio, M. 1982, \apj, 258, 770
\reference{Sato86}Sato, N. Nagase, F., Kawai, N., Kelley, R.L., Rappaport, S.,
	White, N.E. 1986, \apj, 304, 241
\reference{Scw} Schwartz, D.A., Bleach, R.D., Boldt, E.A., Holt. S.S. 
	Serlemitsos, P.J. 1972, \apjl, 173, L51
\reference{Shino}Shinoda, K., {\it et al.} 1990, PASJ, 42, L43
\reference{Soong}Soong, Y. \& Swank, J. H. 1989, in Proc. 23d ESLAB Symp., Two Topics in X-ray
	Astronomy, Vol. 1, ed. N. White (Garching: ESA), 617
\reference{Tak91}Takeshima, T., {\it et al.} 1991, PASJ, 43, L43
\reference{Tak94}Takeshima, T., Dotani, T., Mitsuda, K., \& Nagase, F. 1994, ApJ, 436, 871
\reference{Klis}van der Klis, M., Jansen, F., van Paradijs, J. P., Lewin, W. H. G., Trumper, J.,
	\& Sztajno, M. 1987, ApJ, 313, L19
\reference{Kerkwijk} Van Kerkwijk, M.H., Van Oijen, J.G.J., van den Heuvel, 
	E.P.J. 1989, \aap, 209, 173
\reference{Wang87}Wang, Y.M. 1987, \aap, 183, 257
\reference{Wang95}Wang, Y.M. 1995, \apj, 449, L153
\reference{Zhang}Zhang, W., Giles, A.B., Jahoda, K., Soong, Y. Swank, J.H.,
	Morgan, E.H. 1993, Proc. SPIE, 2006, 324
\reference{Zickgraf}Zickgraf, F.-J., Humphreys, R.M., Lamers, H.J.G.L.M.,
	Smolinski, J., Wolf, B., Stahl, O. 1996, \aap, 315, 510
\end{references}
\end{document}